**Adsorption-Induced Deformation of Mesoporous Solids**

Gennady Yu. Gor* and Alexander V. Neimark#

*Department of Chemical and Biochemical Engineering,*

*Rutgers, The State University of New Jersey,*

*98 Brett Road, Piscataway, New Jersey 08854-8058, United States*

The Derjaguin – Broekhoff – de Boer theory of capillary condensation is employed to describe deformation of mesoporous solids in the course of adsorption-desorption hysteretic cycles. We suggest a thermodynamic model, which relates the mechanical stress induced by adsorbed phase to the adsorption isotherm. Analytical expressions are derived for the dependence of the solvation pressure on the vapor pressure. The proposed method provides a description of non-monotonic hysteretic deformation during capillary condensation without invoking any adjustable parameters. The method is showcased drawing on the examples of literature experimental data on adsorption deformation of porous glass and SBA-15 silica.

* e-mail: ggor@rci.rutgers.edu
# corresponding author, e-mail: aneimark@rutgers.edu

## 1. Introduction

When guest molecules are adsorbed in pores of a solid body, they induce elastic stress in the host matrix and cause its deformation[1]. This phenomenon of adsorption-induced deformation is ubiquitous, although its magnitude in many systems is negligibly small. The detailed experimental works date back to 1920s when Meehan[2] and Bangham and Fakhoury[3] published their dilatometric studies on expansion of charcoal upon adsorption of $CO_2$ and other vapors and gases. Bangham et al.[4, 5] put forward the first theory of adsorption deformation relating the progressive matrix expansion to the gradual decrease of the surface free energy in the course of adsorption that is referred to as the Bangham effect.

Later, adsorption deformation was studied in various microporous and mesoporous solids [1, 6-26]. It was found that adsorption deformation displays qualitatively different behaviors in different systems. Adsorption-induced expansion due to the Bangham effect is not universal. Many microporous solids, including certain active carbons and zeolites, exhibit in the course of adsorption a non-monotonic deformation, contracting at low vapor pressures and expanding at higher pressures[6]. Some mesoporous solids, like porous glasses and gels may demonstrate several alternating stages of expansion and contraction, which are associated with different adsorption mechanisms[9, 15, 22]. Therewith, the adsorption hysteresis typical for mesoporous materials gives rise to the deformation hysteresis during adsorption-desorption cycles.

Recently, the interest in adsorption deformation was re-ignited by the development of novel nanoporous materials and new experimental techniques of ellipsometric porosimetry[27] and *in situ* small angle X-ray and neutron diffractometry[28]. These studies, especially recent *in situ* XRD measurements with well characterized MCM-41 and SBA-15 mesoporous crystals[13, 14] and metal organic frameworks[29], provide the most reliable experimental data that can be employed for verification of theoretical models.

After the works of Bangham et al.[4, 5], there were several attempts to build a thermodynamic theory of adsorption deformation[16,8, 30] based on the ideal solution theory of adsorption in microporous solids. As related to mesoporous solids, the deformation is beyond the Bangham effect. In their seminal work Amberg and McIntosh[9] suggested a qualitative description of the competition between the Bangham and capillary effects in the course of adsorption-desorption hysteresis of water on porous glass. Experimental data[9] served as a challenging case study for Grosman and Ortega[20, 31], who considered adsorption deformation of porous silicon. We will also use this data below to illustrate the validity of the proposed method. Eriksson[32] presented a thermodynamic analysis of adsorption strain taking into account the specific properties of a solid surface. Unlike Bangham effect, Eriksson's theory explained not only sample expansion, but contraction also for the case of localized adsorption of molecules with lateral interactions. Ash, Everett, and Radke [33] suggested an alternative thermodynamic formulation to consider a competition between expansion due to the Bangham effect and contraction due to dispersion forces. Scherer[34] reexamined the Bangham and Yates theories with example of adsorption deformation of porous glass and suggested a relationship between the elastic modulus and the Young modulus and Poisson ratio of the solid.

Several original theoretical approaches to adsorption deformation problems were suggested recently. Rusanov and Kuni[18, 19] calculated the stress of the single pore using the Irving-Kirkwood expression for the pressure tensor[35] and obtained a body strain using the macroscopic theory of elasticity[36]. Mushrif and Rey[24] attempted to apply the Biot theory of poroelasticity[37, 38] to calculate the difference of chemical potentials of fluids adsorbed on strained and unstrained substrates as a function of the porosity change. Specially designed Monte Carlo simulations methods were proposed to model adsorption in deformable pores, which employed statistico-mechanical formalisms of the osmotic ensemble[39] and the semi-grand canonical ensemble[40].

In this paper, we extend to mesoporous materials a simple yet instructive thermodynamic model of the adsorption stress, which was successfully employed in the previous works on adsorption deformation of microporous solids like zeolites[17], active carbons[21], and MOFs[26]. The proposed model relates the stress, exerted by the adsorbed phase on the adsorbent framework, with the adsorption isotherm. From the thermodynamic standpoint, the adsorption stress $\sigma_S$ can be quantified by the derivative of the grand thermodynamic potential $\Omega$ of the adsorbed phase with respect to the pore volume $V$ at fixed temperature $T$ and adsorbate chemical potential $\mu$[17]

$$\sigma_s = -\left(\frac{\partial \Omega}{\partial V}\right)_{\mu,T} . \tag{1}$$

In pores of simple geometries of slit, cylindrical, or spherical shapes, the adsorption stress has a simple physical interpretation as a normal to the pore wall component of the stress tensor in the adsorbed phase [17, 21, 41]. In disordered materials this interpretation is no longer valid, since the normal stress tensor component depends on the pore wall curvature and varies in space. However, the adsorption stress defined by Eq. 1 can serve as an overall scalar measure of the magnitude of the adsorption forces acting on the porous framework. The advantage of the thermodynamic model based on Eq. 1 is its universality: it relates the mechanical properties to the variation of the grand thermodynamic potential that can be determined by using suitable statistico-mechanical, molecular simulation, or empirical methods.

The difference between the adsorption stress, $\sigma_S$, and the external pressure $P$ represents the so-called solvation pressure $f_S$, which determines the magnitude of elastic deformation in terms of the volumetric strain $\varepsilon$ ($\varepsilon = \Delta V/V$, where $\Delta V$ is the variation of the cell volume), assuming the linear Hooke law with an effective elastic modulus $K$,

$$f_s = \sigma_s - P = K\varepsilon + \sigma_0 , \tag{2}$$

where $\sigma_0$ is a pre-stress in the reference state at which the initial sample volume $V$ is defined.

The linear elasticity theory describes adsorption-induced deformation of microporous and mesoporous materials like zeolites, activated carbons, porous glasses, and mesoporous crystals, when the strain is small, typically in fractions of percent. Eq. 2 was employed for the determination of the elastic modulus $K$ from dilatometric, ellipsometric, and XRD measurements of adsorbent deformation in the region of capillary condensation[9, 13, 23, 25, 42]. These authors implied that the stress in the RHS of Eq. 2 equals to the Laplace pressure acting on liquid-vapor interfaces, namely,

$$f_s = \mu / V_L = \left(R_g T / V_L\right) \ln(P/P_0)\,, \qquad (3)$$

where $R_g$ is the gas constant, $V_L$ is the molar volume of the condensed phase and $P_0$ is the pressure of saturated vapor. Eq. 3 implies that vapor is ideal gas and condensed phase is incompressible liquids. In the following discussion Eq. 3 is referred to as the capillary approximation. As shown in[13, 42] the capillary approximation provides a reasonable quantitative description of the initial adsorbent contraction of various mesoporous solids in the process of desorption/drying from the fully saturated state. However, the capillary approximation does not take into account the adsorbate-solid interactions and, thus, cannot explain a diversity of non-monotonic strain isotherms observed for different mesoporous solids.

To analyze adsorption deformation of mesoporous materials, we employ below the Derjaguin theory of thin film equilibrium[43], which was implemented for quantitative description of capillary condensation-desorption cycles in mesoporous solids by Broekhoff and de Boer[44, 45]. Although the Derjaguin – Broekhoff – de Boer (DBdB) theory is purely macroscopic, and it cannot be expected to be adequate in small pores comparable in size with molecular diameters, it was shown to provide a correct qualitative description of the adsorption-desorption process in pores of simple geometry such as cylindrical and spherical pores[46, 47]. We demonstrate that the DBdB theory can be used for determining in a unified fashion the variation of the adsorption stress (Eq. 1) within the adsorption process starting from the formation and growth of adsorption films to capillary condensation. The proposed method provides an analytical description of non-monotonic hysteretic deformation during capillary condensation-desorption cycles without invoking any adjustable parameters. We showcase this approach drawing on the examples of classical experiments of Amberg and McIntosh [9] and most recent experimental data, obtained using *in situ* small-angle X-rays scattering on SBA-15 silica[14].

## 2. DBdB theory of capillary condensation

The capillary approximation (3) ignores the solid-fluid interactions, which are responsible for the Bangham effect. In the following discussion, we employ the formalism of the disjoining pressure that was introduced by Derjaguin[43] to account for solid-fluid interactions. According to the DBdB theory, the adsorption process is treated as formation of adsorption films on curved wetting pore walls. Equilibrium and stability of adsorption films are determined by a balance of capillary and disjoining pressures. For a cylindrical pore of radius *R*, the chemical potential $\mu$ of adsorption film of thickness *h* in equilibrium with vapor at pressure *P* and temperature *T* is given by the Derjaguin equation,

$$\mu = R_g T \ln(P/P_0) = -\left(\Pi(h) + \frac{\gamma}{R-h}\right)V_L, \tag{4}$$

where $\Pi(h)$ is the disjoining pressure, $\gamma$ is the liquid-vapor surface tension. Eq. 4 represents an algebraic equation for the film thickness as a function of relative pressure, which can be easily transformed into the equation for adsorption isotherm. Eq. 4 treats gas phase as ideal and condensed phase and adsorbed film as incompressible liquid.

As a typical example, we present in Fig. 1a the DBdB adsorption-desorption isotherm of nitrogen at *77.4 K* in cylindrical pore of *R = 4.1 nm*, which was calculated using the Frenkel-Halsey-Hill (FHH) form of the disjoining pressure[48],

$$\Pi(h) = \frac{R_g T}{V_L}\frac{k}{(h/h_0)^m}, \tag{5}$$

with $h_0 = 1Å$ and empirical parameters *k =44.54* and *m=2.241* recommended for nitrogen adsorption on silicas[49], we used $\gamma = 8.88\, mN/m$ and $V_L = 3.466\times10^{-6}\ m^3/mol$ [46]. This example was chosen to approximate the experimental isotherm of nitrogen adsorption on the sample of SBA-15 silica[50], which was used in *in situ* studies of adsorption deformation in reference[14], discussed below. To take into account the intrawall microporosity inherent to SBA-15 materials, the DBdB isotherm was shifted upwards to match the theoretical and experimental points E.

The adsorption isotherm (4) is depicted by a solid line OECFS in Fig. 1a. This isotherm represents the thickness of the adsorbed film forming on the pore walls as the vapor pressure increases. The theoretical isotherm matches nicely with the experimental adsorption isotherm up to the point of capillary condensation, which corresponds to the turnover spinodal point C. The spinodal determines the limit of film stability and reflects the conditions of spontaneous capillary condensation depicted by the vertical line CF in Fig. 1a. The condition of condensation is the maximum of the film chemical potential given by Eq. 4. The critical film thickness $h_c$ at point C is determined from the following algebraic equation

$$\frac{d\Pi(h)}{dh}+\frac{\gamma}{(R-h)^2}=0\,. \tag{6}$$

Vapor pressure $P_c$ of capillary condensation is given by Eq. 4 at $h=h_c$.

The process of desorption from completely filled pores (point S in Fig. 1a) is associated with formation of concave menisci of increasing curvature at the pore entrances. In the course of desorption along the initial stage SD, the magnitude of the negative capillary pressure acting across the vapor-liquid interface increases according to Eq. 3 which causes stretching of the condensed liquid and contraction of the solid framework. The desorption point D corresponds to the formation of the equilibrium meniscus, which recedes gradually, transiting into a film of thickness $h_e$ (segment DE in Fig. 1a. The condition of film-meniscus equilibrium for cylindrical pores is determined by Derjaguin equation [44]

$$R_g T\ln(P/P_0)=-2V_L\left[\frac{\gamma}{R-h}+\frac{1}{(R-h)^2}\int_h^R (R-h')\Pi(h')dh'\right]. \tag{7}$$

Eqs. 4 and 7 determine the equilibrium film thickness $h_e$ and corresponding vapor pressure $P_e$ of desorption.

The DBdB theory qualitatively explains the origin of capillary hysteresis in open-ended pore channels and, as seen from Fig. 1a, agrees reasonably well with experimental results. It is worth noting that the DBdB theory is a macroscopic approach, which cannot be justified for small pores comparable in size with molecular dimensions. The estimates[46, 47, 51] show that DBdB predictions progressively deviate from a more precise density functional theory (DFT) in pores smaller than $\sim 7\,nm$ with the lower limit of applicability of ~ 4 *nm*.

**3. Adsorption stress calculation**

To analyze the adsorption-induced deformation, we have to calculate the adsorption stress $\sigma_s$ defined by Eq. 1, which for cylindrical pore geometry reduces to

$$\sigma_s = -\frac{1}{2\pi RL}\left(\frac{\partial\Omega}{\partial R}\right)_{\mu,T}, \tag{8}$$

where $L$ is the pore length. Let us consider the variation of the adsorption stress and respectively the solvation pressure along the adsorption-desorption cycle as a function of the vapor pressure, which varies from $P = 0$ (dry pore) to $P = P_0$ (saturation). In so doing, we will perform calculations for three characteristic adsorption regimes: complete pore filing ($P_e \le P \le P_0$ for desorption branch and $P_c \le P \le P_0$ for adsorption branch), adsorption film region ($0 < P \le P_e$ for desorption branch and $0 < P \le P_c$ for adsorption branch), and capillary condensation and evaporation transitions at $P = P_c$ and $P = P_e$. As an instructive example, we present in Fig. 1b the results of solvation pressure calculations for the isotherm of nitrogen adsorption on SBA-15 presented in Fig. 1a. Since the FHH equation (Eq. 5) employed in calculations becomes progressively inaccurate at low pressures, the solvation pressure isotherm is given only for $P / P_0 > 0.05$. In Fig. 1b, we present the change of the solvation pressure compared to the saturation state, $f_s + \gamma_{sl} / R$ ($\gamma_{sl}$ is the pore wall – liquid surface tension), which corresponds to the choice of the saturation state as the reference state to reckon the sample strain.

For a dry pore ($P = 0$), the grand thermodynamic potential is equal to

$$\Omega(0) = 2\pi RL\gamma_s , \tag{9}$$

where $\gamma_s$ is the pore wall surface tension at $P \to 0$ (dry solid). With respect to Eq. 8, the stress in the evacuated sample in the beginning of adsorption is determined by the Laplace equation for a cylindrical interface,

$$\sigma_s(0) = \sigma_0 = -\gamma_s / R . \tag{10}$$

This value determines the limiting solvation pressure at $P \to 0$.

For the saturation state of complete pore filling at $P = P_0$, we have

$$\Omega(P_0) = -\pi R^2 L P_0 + 2\pi RL\gamma_{sl} , \tag{11}$$

provided that the pore is sufficiently long ($L >> R$), so that the contributions from the entrance effects are negligible. Adsorption stress for the saturation state is equal to the sum of capillary pressure on cylindrical wall – liquid interface and the external pressure of saturated vapor

$$\sigma_s(P_0) = -\gamma_{sl} / R + P_0 . \tag{12}$$

The magnitude of sample deformation during either adsorption or desorption process is determined according to Eq. 2 by the solvation pressure difference between the saturated and dry states,

$$\Delta f_s = \sigma_s(P_0) - P_0 - \sigma_0 = -(\gamma_{sl} - \gamma_s) / R . \tag{13}$$

The variation of the grand thermodynamic potential along the isotherm is determined by the Gibbs equation,

$$\left(\frac{\partial \Omega}{\partial \mu}\right)_{T,V} = -N(\mu) , \tag{14}$$

where $N$ is a number of adsorbate molecules in the pore . Within the pore filling region variation of the grand thermodynamic potential is given by integration of Gibbs equation (14)

$$\Omega(P) = \Omega(P_0) - \int_0^{\mu(P)} N(\mu')d\mu' = 2\pi RL\gamma_{sl} - \pi R^2 L P_0 - \frac{\pi R^2 L}{V_L}\mu(P) . \tag{15}$$

Eq. 15 implies that within the DBdB approximation condensed liquid is incompressible, and thus,

$N = \pi R^2 L / V_L$. Eqs. 8 and 15 with Eq. 4 for the chemical potential of ideal vapor immediately allow us to recover the capillary approximation for the solvation pressure $f_s$ within the pore filling region,

$$f_s(P) = -\frac{\gamma_{sl}}{R} + \frac{R_g T}{V_L} \ln\left(\frac{P}{P_0}\right) + (P_0 - P)\,. \tag{16}$$

The term ( $P_0 - P$ ) accounts for the change of the external pressure, which is usually small compared with the capillary term and can be neglected for practical estimates. However for the case of supercritical adsorption at high pressure (which is beyond the scope of the current paper) this term should be kept. In Fig. 1b, Eq. 14 is depicted as SD section for desorption and FS section for adsorption.

For the film region, the grand thermodynamic potential is determined by integration of the Gibbs equation (14) from the dry state,

$$\Omega^f(P) = \Omega(0) - \int_{-\infty}^{\mu(P)} N^f(\mu') d\mu'\,, \tag{17}$$

where $N^f = \frac{\pi L}{V_L}\left[R^2 - (R-h)^2\right]$ is a number of adsorbate molecules in the film. By differentiating Eq. 17 with respect to R (see supplementary materials for detailed derivation), we obtain the relationship between the adsorption stress and the equilibrium film thickness for given vapor pressure and temperature,

$$\sigma_s{}^f(P) = -\frac{\gamma_s - \int_{\Pi(h)}^{\infty} h' d\Pi(h')}{R} - \left[\frac{\gamma}{R-h} - \frac{\gamma}{R}\right] = -\frac{\gamma_s - \gamma}{R} - \frac{\gamma}{R-h} - \frac{h}{R}\Pi(h) + \frac{1}{R}\int_0^h \Pi(h') dh'\,. \tag{18}$$

Noteworthy, as seen from the example in the Inset in Fig. 1 (curve EC), this dependence may be non-monotonic due to a competition between the contributions from capillary and disjoining pressures. The first term in the RHS of Eq. 18, which is monotonically increasing for wetting surfaces from the dry state value of $\sigma_s(0) = \sigma_0 = -\gamma_s / R$, reflects the Bangham effect of reducing the surface energy in the process of adsorption that is associated with the sample expansion. Indeed, $\gamma_s - \int_{\Pi(h)}^{\infty} h' d\Pi(h')$ represents the free energy of the flat surface covered by the adsorption film of thickness $h$. The capillary pressure term is always negative; it decreases from 0 at the dry state. Thus, the adsorption stress first increases due to the Bangham effect and then once the magnitude of capillary pressure becomes comparable with the effects of disjoining pressure, it achieves a maximum and decreases, see Inset in Fig. 1b. The position of maximum stress (point M in the Inset in Fig. 1b) is determined by the following condition,

$$\frac{h}{R}\frac{d\Pi(h)}{dh} + \frac{\gamma}{(R-h)^2} = 0. \qquad (19)$$

As seen from the plot the maximum stress is achieved in the region between the equilibrium and capillary condensation pressures. Moreover, from the condition (6), it follows that $\sigma_s(P/P_0)$ curve has a vertical tangent at $P \to P_c$. Thus, we may conclude that the adsorption stress is non-monotonic in the process of adsorption prior to capillary condensation and monotonically decreasing in the process of desorption. This behavior was experimentally shown by Amberg and McIntosh[9], see below Fig. 3b.

While Eqs. 16 and 18 show how the solvation pressure varies within the pore filling and film regions, the stepwise increase of the solvation pressure (segment ED in Fig. 1b) at the desorption point and its drop at the condensation point (segment FC in Fig. 1b) are given by the following relationships:

$$\Delta f_c \equiv \sigma_s{}^f(P_c) - \sigma_s(P_c) = -\frac{\gamma_s - \gamma_{sl} - \gamma}{R} - P_0 + \Pi(h_c)\left(1 - \frac{h_c}{R}\right) + \frac{1}{R}\int_0^{h_c} \Pi(h)dh \qquad (20)$$

and

$$\Delta f_e \equiv \sigma_s{}^f(P_e) - \sigma_s(P_e) = -\frac{\gamma_s - \gamma_{sl} - \gamma}{R} - P_0 + \Pi(h_e)\left(1 - \frac{h_e}{R}\right) + \frac{1}{R}\int_0^{h_e} \Pi(h)dh \qquad (21)$$

Eqs. 18, 20 and 21 can be complemented by the Frumkin-Derjaguin equation, which relates the liquid-vapor, solid-liquid, and dry solid surface tensions with the disjoining pressure isotherm, (see e.g. Ref.

[52]),

$$\int_0^\infty \Pi(h)dh = \gamma_s - \gamma_{sl} - \gamma, \tag{22}$$

Eq. 22 allows us to reduce the number of physical parameters that should be used for calculating solvation pressure during the adsorption-desorption cycle, and eliminate the dry solid surface tension $\gamma_s$ that cannot be readily measured. Indeed, Eqs. 18 and 20, 21 are transformed into

$$\sigma_s{}^f(P) = -\frac{\gamma_{sl}}{R} - \frac{\gamma}{R-h} - \frac{h}{R}\Pi(h) - \frac{1}{R}\int_h^\infty \Pi(h')dh', \tag{23}$$

$$\Delta f_c = -P_0 + \Pi(h_c)\left(1 - \frac{h_c}{R}\right) - \frac{1}{R}\int_{h_c}^\infty \Pi(h')dh', \tag{24}$$

$$\Delta f_e = -P_0 + \Pi(h_e)\left(1 - \frac{h_e}{R}\right) - \frac{1}{R}\int_{h_e}^\infty \Pi(h')dh'. \tag{25}$$

All parameters in Eqs. 23-25, the solid-liquid and liquid-vapor surface tensions, and the disjoining isotherm in the region of thick films can be determined by adsorption and wetting experiments on non-porous surface. Moreover, the magnitudes of the solvation pressure steps at the point of condensation and desorption transitions, $\Delta f_c$ (24) and $\Delta f_e$ (25), do not depend on the solid-liquid surface tension, and are determined by the disjoining pressure isotherm in the region $h > h_e$. For the FHH isotherm (5), Eq. 23 for solvation pressure reads

$$f_s{}^f(h) = -\frac{\gamma_{sl}}{R} - P - \frac{\gamma}{R-h} - \frac{R_g T}{V_L} k \frac{m}{m-1} \frac{h_0}{R}\left(\frac{h}{h_0}\right)^{1-m}. \tag{26}$$

Eq. 26 was employed in calculations presented in Fig.1b.

**4. Comparison with experimental data**

The results of calculations for nitrogen adsorption of SBA-15 given in Fig.1b, shows the specifics of the deformation process, which can be confirmed by experiment. Unfortunately, relevant data for nitrogen adsorption are not available. However, the authors [14] reported *in situ* XRD data on deformation of the same sample in the course of water desorption at temperature *T = 293 K*, which we employed for comparison with the theoretical predictions. For calculation of the solvation pressure of adsorbed water, we used the pore radius $R \approx 4.1\,nm$ determined from the nitrogen adsorption, and the parameters of FHH isotherm of water on silica $k = 9.5$ and $m = 2$, reported in Ref.[53], $\gamma = 7.2\times10^{-2}\ N/m$, $V_L = 1.8\times10^{-5}\ m^3/mol$.

In Fig. 2, we compare the calculation results with experimental data on water desorption. In order to make such comparison we convert the solvation pressure $f_s$ into the strain $\varepsilon$ using the elastic modulus $K = 6\,GPa$ reported in Ref. [14]. Since the experimental strain is reckoned from the dry state and the values of the dry solid and solid-liquid surface tensions are not known, the theoretical and experimental strain isotherms have been matched by a vertical shift at the position of the onset of capillary desorption at $P_e/P_0 \approx 0.69$, which corresponds the point D of the largest contraction. Agreement between the theory and experiment is almost quantitative: the magnitudes of contraction during desorption prior to capillary evaporation and expansion during capillary evaporation, as well as the strain variation during film desorption coincide fairly well. An apparent deviation of the shape of the experimental strain isotherm during capillary evaporation (tilted step instead of vertical) can be attributed to an inherent polydispersity of pore sizes and, possibly, to an insufficient equilibration time during experiments. The maximum strain $\varepsilon_s$ at the saturation gives us according to Eq. 13 an estimate of the difference between the dry solid and solid-liquid surface tension, $\gamma_s - \gamma_{sl} = KR\varepsilon_s \approx 0.13\ N/m$.

In Fig. 3, we attempted to compare the proposed theory with experiments of Amberg and McIntosh on water adsorption on porous glass [9]. Due to the lack of relevant experimental data, we used the same FHH disjoining pressure isotherm of water on silica employed above for water/SBA-15 system. This isotherm provides a reasonable qualitative description of the adsorption-desorption hysteresis loop, as seen in Fig. 3a, where we plot the theoretical adsorption-desorption isotherms as in a cylindrical pore of $R = 3.95\,nm$. This pore size was used for calculating the solvation pressure, which was converted into the strain employing Eq. 2 with the elastic modulus $K = 38\ GPa$ reported in Ref. [9]. Since we use a single pore model and the incompressible liquid approximation, the calculated desorption branch prior to capillary evaporation (section FD in Fig. 1a) is flat. Due to this reason, we normalized the adsorption isotherm by the adsorption amount at the upper closure point of the hysteresis loop, at $P/P_0 \approx 0.77$. The calculated and experimental isotherms matched at the onset of capillary evaporation (point D), similarly to the previous example. Note that the proposed theory correctly predicts the shape of the hysteresis loops in both adsorption and stress measurements. An apparent deviation in the film region can be attributed to the fact that FHH disjoining pressure isotherm used in calculations poorly describes water adsorption on glass surface at low coverages.

## 5. Conclusions

The Derjaguin – Broekhoff – de Boer theory of capillary condensation was employed to describe deformation of mesoporous solids in the course of adsorption-desorption hysteretic cycles. We suggested a thermodynamic model, which relates the mechanical stress induced by adsorbed phase with the adsorption isotherm. Analytical expressions were derived for the dependence of the solvation pressure on the vapor pressure. The proposed method allows one to predict the variation of the solvation pressure and, respectively, the sample deformation based on independent experimental information. The method was verified with experimental data obtained recently by *in situ* XRD measurements on SBA-15 mesoporous crystals[14], as well as with classical data of dilatometric measurements on porous glass [9]. The proposed method is in line with the qualitative description of Amberg and McIntosh[9] of a competition of Bangham and capillarity effects, and it provides a quantitative account of this phenomenon.

The proposed method can be implemented for other mesoporous materials assuming pores of different geometry (e. g. spherical, cylindrical, or slit-shaped). It can be generalized by taking into account the pore-size distribution according to the ideas outlined earlier in Ref. [21] It is worth noting, that from the comparison of calculated and experimental strain isotherm, one can estimate the elastic modulus of saturated and partially saturated porous bodies.

The limits of applicability of the proposed method are related with the limitations of DBdB theory. As shown in Refs. [46, 47, 51], while DBdB theory provides a reasonable quantitative description of capillary condensation phenomenon in pores larger than *7-8 nm*, its accuracy progressively diminishes as the pore size decreases. In smaller pores, the molecular level methods, such as Monte Carlo simulation and density functional theory should be employed for determining theoretical adsorption isotherms and thermodynamic potentials required for adsorption stress calculations.

**Acknowledgements**

This work was supported in parts by the NSF ERC “Structured Organic Particulate Systems” and the Blaise Pascal International Research Chair fellowship to A.V.N. The authors thank George Scherer, Christopher Rasmussen, and Aleksey Vishnyakov for helpful comments.

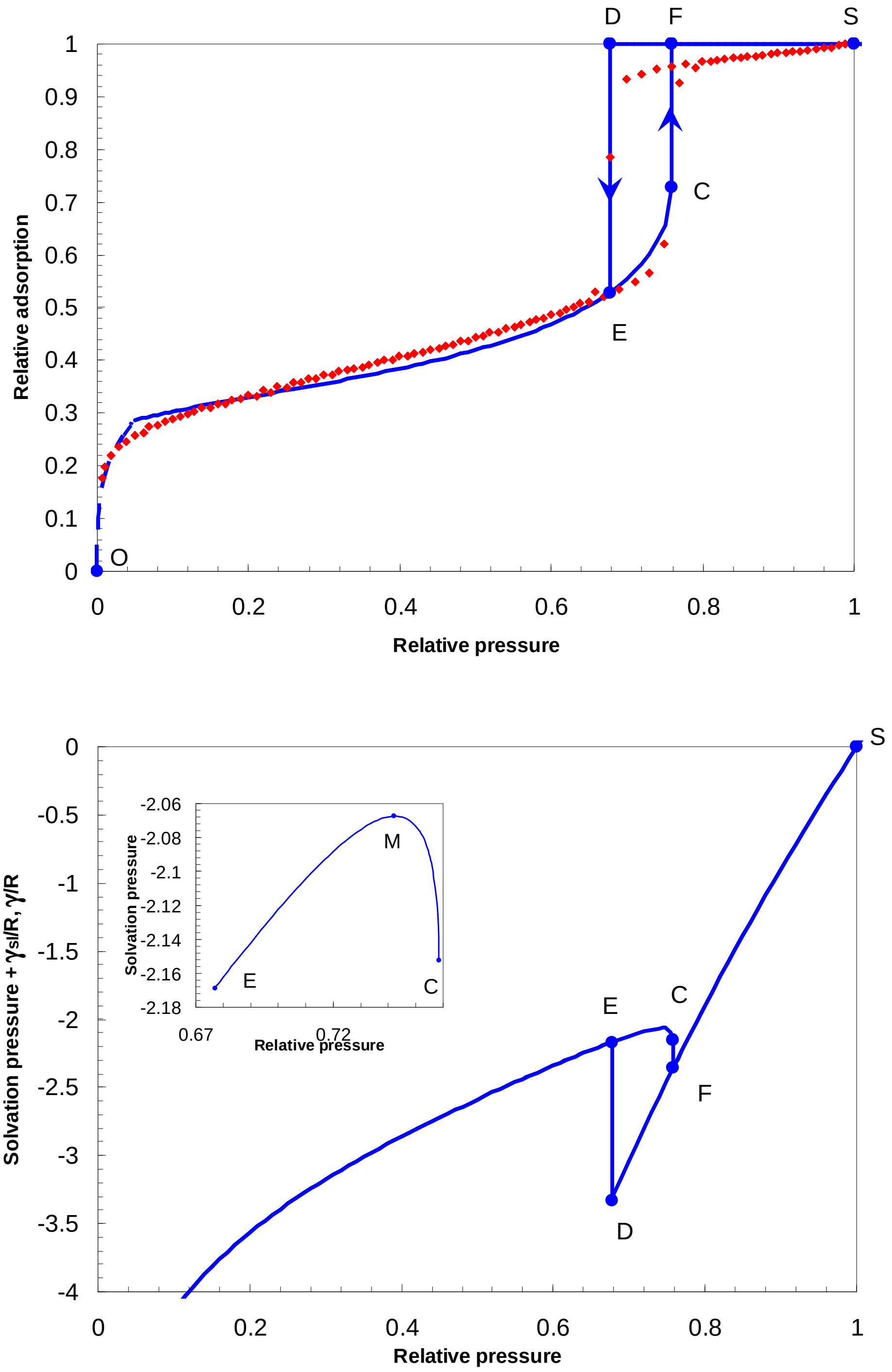

**Fig. 1.** Nitrogen adsorption and desorption on SBA-15 silica at *77.4 K* [50]. **(a)** Adsorption-desorption isotherms reduced to the maximum adsorption at saturation. Points – experimental data[50]. Theoretical isotherm was calculated in *8.2 nm* pore using Eqs. 4 and 5 and shifted upward to account for the sample microporosity by matching the lower closure point E of the hysteresis loop. Continuation of the theoretical isotherm at low pressures are marked by dash line that should be treated as a guide for the eye. **(b)** Solvation pressure in $\gamma/R$ units calculated using Eq. 16 for DFS branch and Eq. 26 for EC branch. The solvation pressure is shifted by $\gamma_{sl}$/R. This shift corresponds to the choice of the saturation state at $P/P_0 = 1$ as the reference state for reckoning the strain. **(Inset)** Magnified section EC from Fig. 1b shows the maximum expansion (point M) achieved in the course of adsorption prior to capillary condensation.

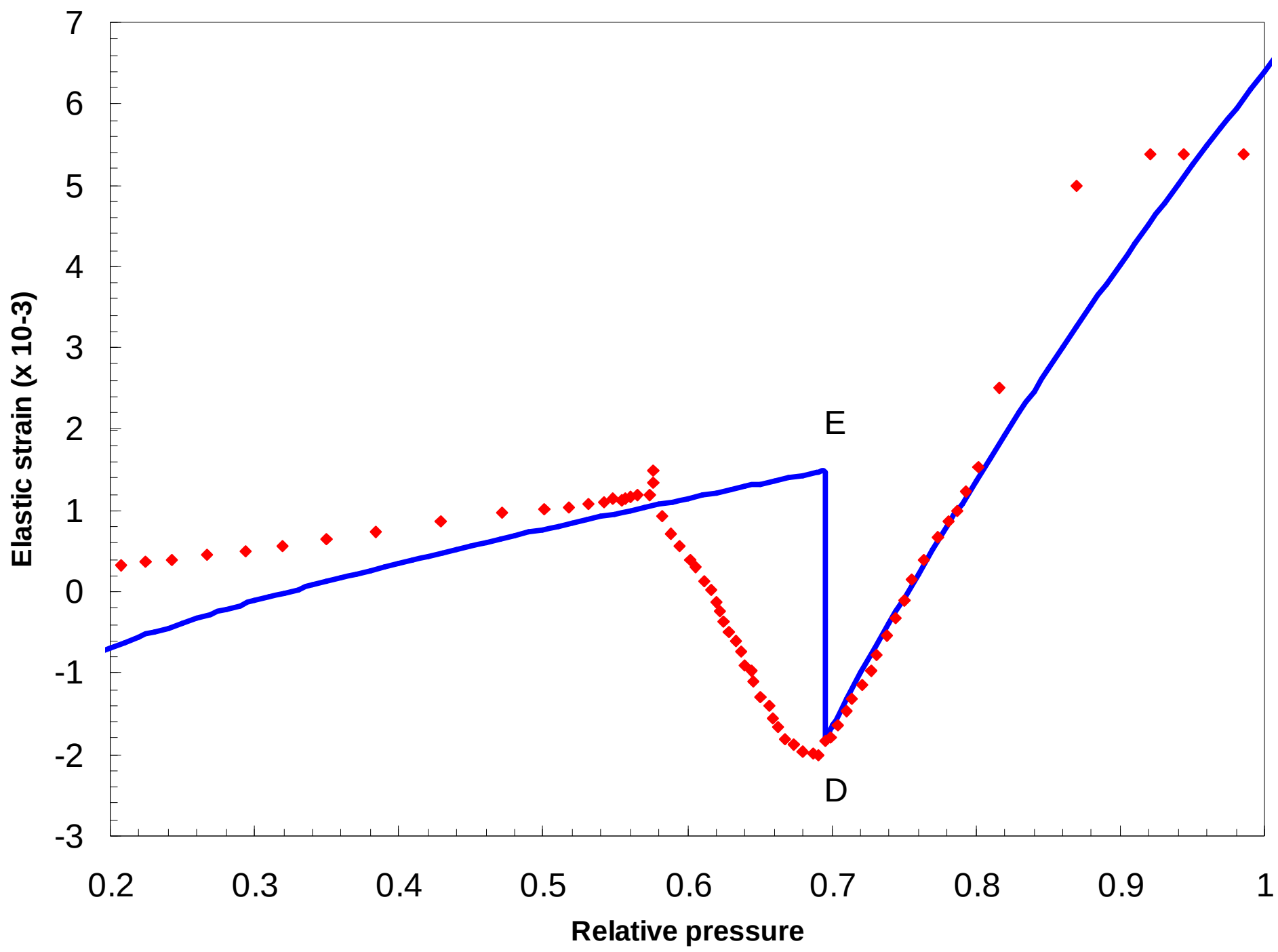

**Fig. 2.** Water desorption on SBA-15 silica [14] at 293 K. Sample strain $\varepsilon$ as a function of relative pressure $P/P_0$. Points – experimental data [14], solid curve – calculations in *8.2 nm* pore according to Eq. 16 (part of the curve to the right of point D) and Eq. 26 (part of the curve to the left of point E). The strain is determined from the solvation pressure using the bulk modulus $K = 6\,GPa$ [14] in Eq. 2. Experimental and theoretical data are matched by a vertical shift at the point D of the largest contraction. Note, the experimental strain was reckoned from the dry state at $P/P_0 \to 0$. An apparent difference with the presentation in Fig. 1b is due to the different reference states (saturation state in Fig. 1b and dry state in Fig. 2).

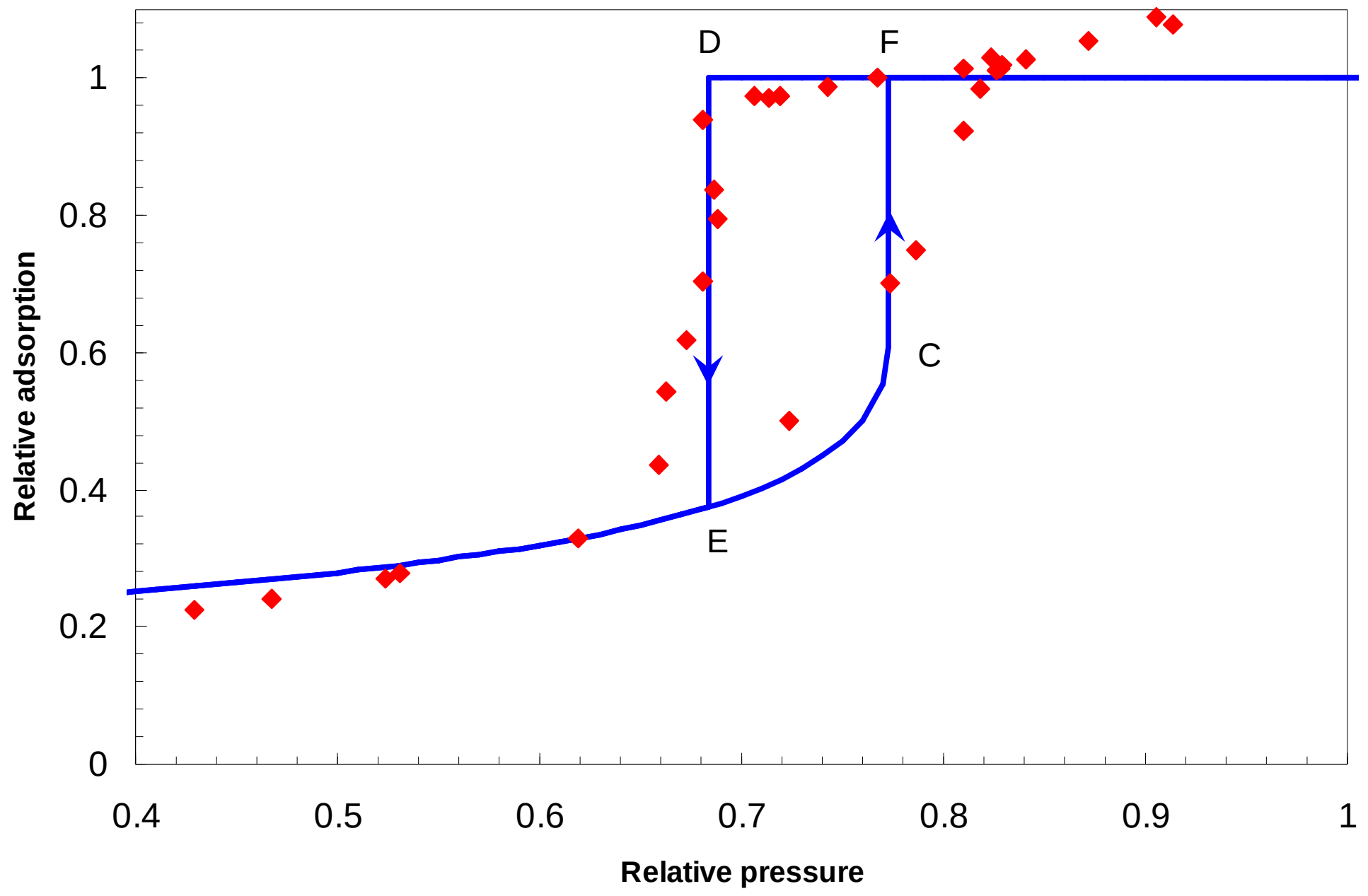

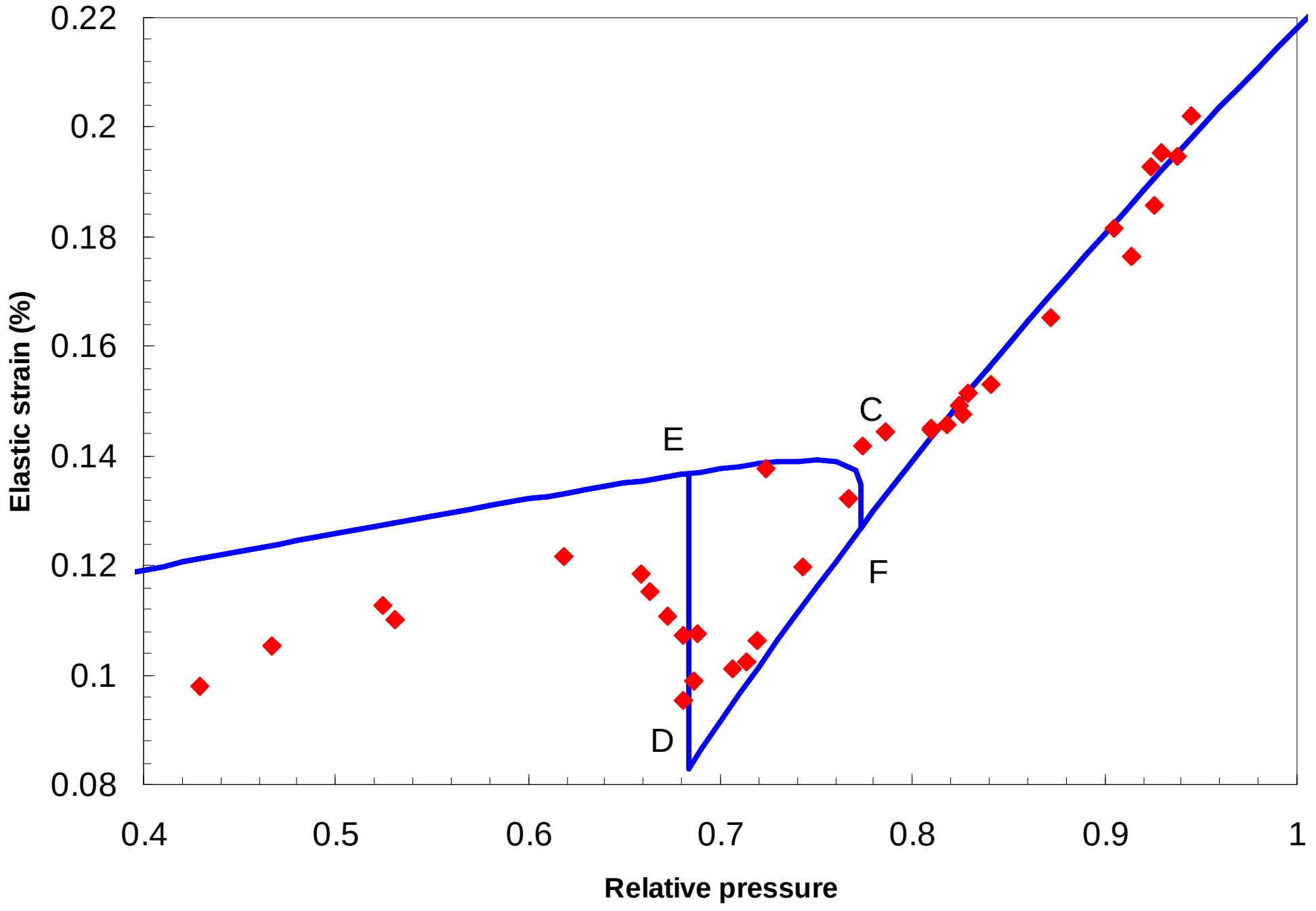

**Fig. 3.** Water adsorption and desorption on porous glass[9] at 291.9 K. **(a)** Adsorption-desorption isotherms as a function of pressure $P/P_0$ (points – experimental data [9], lines – calculated in *7.9 nm* pore with Eqs. 4 and 5). The experimental and theoretical isotherms are normalized by the adsorption amount at the upper closure point of the hysteresis loop and matched at the lower closure point similarly to Fig. 1a. **(b)** Sample strain $\varepsilon$ as a function of relative pressure $P/P_0$. Points – experimental data [9], lines – calculations according to Eqs. 16 and 26. The strain is determined from the solvation pressure using the bulk modulus $K = 38\ GPa$ [9] in Eq. 2. Note, the experimental strain was reckoned from the dry state at $P/P_0 \to 0$, and experimental and theoretical data are matched by a vertical shift at the point D of the largest contraction similarly to Fig. 2.

## References

1.Tvardovskiy, A. V., *Interface Science and Technology - V. 13, Sorbent deformation*. Elsevier Ltd.: Oxford, 2007; p 286.

2.Meehan, F. T., The Expansion of Charcoal on Sorption of Carbon Dioxide. *Proc. R. Soc. Lond. A* **1927,** 115, 199-207.

3.Bangham, D. H.; Fakhoury, N., The expansion of charcoal accompanying sorption of gases and vapours. *Nature* **1928,** 122, 681-682.

4.Bangham, D. H.; Fakhoury, N.; Mohamed, A. F., The Swelling of Charcoal. Part II. Some Factors Controlling the Expansion Caused by Water, Benzene and Pyridine Vapours. *Proceedings of the Royal Society of London. Series A, Containing Papers of a Mathematical and Physical Character* **1932,** 138, (834), 162-183.

5.Bangham, D. H.; Fakhoury, N.; Mohamed, A. F., The Swelling of Charcoal. Part III. Experiments with the Lower Alcohols. *Proc. R. Soc. Lond. A* **1934,** 147, 152-175.

6.Haines, R. S.; McIntosh, R., Length Changes of Activated Carbon Rods Caused by Adsorption of Vapors. *The Journal of Chemical Physics* **1947,** 15, (1), 28-38.

7.Yates, D. J. C., The Expansion of Porous Glass on the Adsorption of Non-Polar Gases. *Proceedings of the Royal Society of London Series A-Mathematical and Physical Sciences* **1954,** 224, (1159), 526-544.

8.Bering, B. P.; Krasilnikova, O. K.; Serpinskii, V. V., Deformation of Zeolite CaA Granule During Krypton Adsorption. *Bulletin of the Academy of Sciences of the USSR Division of Chemical Science* **1978,** 27, (12), 2515-2517.

9.Amberg, C. H.; McIntosh, R., A study of adsorption hysteresis by means of length changes of a rod of porous glass. *Canadian Journal of Chemistry* **1952,** 30, 1012.

10.Fomkin, A. A., Adsorption of gases, vapors and liquids by microporous adsorbents. *Adsorption-Journal of the International Adsorption Society* **2005,** 11, (3-4), 425-436.

11.Scherer, G. W., Drying Gels .1. General-Theory. *Journal of Non-Crystalline Solids* **1986,** 87, (1-2), 199-225.

12.Shkolin, A. V.; Fomkin, A. A., Deformation of AUK Microporous Carbon Adsorbent Induced by Methane Adsorption. *Colloid journal* **2009,** 71, (1), 119-124.

13.Guenther, G.; Prass, J.; Paris, O.; Schoen, M., Novel insights into nanopore deformation caused by capillary condensation. *Physical review letters* **2008,** 101, (8), 086104

14.Prass, J.; Muter, D.; Fratzl, P.; Paris, O., Capillarity-driven deformation of ordered nanoporous silica. *Applied physics letters* **2009,** 95, (8), 083121.

15.Smith, D. M.; Scherer, G. W.; Anderson, J. M., Shrinkage During Drying of Silica-Gel. *Journal of Non-Crystalline Solids* **1995,** 188, (3), 191-206.

16.Jakubov, T. S.; Mainwaring, D. E., Adsorption-induced dimensional changes of solids. *Physical Chemistry Chemical Physics* **2002,** 4, (22), 5678-5682.

17.Ravikovitch, P. I.; Neimark, A. V., Density functional theory model of adsorption deformation. *Langmuir* **2006,** 22, (26), 10864-10868.

18.Rusanov, A. I., Chemomechanical effects in nanoporous bodies. *Doklady physical chemistry* **2006,** 406, 49-52.

19.Rusanov, A. I.; Kuni, F. M., On the theory of the mechanochemical sorption-striction phenomenon in nanoporous bodies with dispersion forces. *Russian journal of general chemistry* **2007,** 77, (3), 371-392.

20.Grosman, A.; Ortega, C., Influence of elastic deformation of porous materials in adsorption-desorption process: A thermodynamic approach. *Physical review B* **2008,** 78, (8), 085433

21.Kowalczyk, P.; Ciach, A.; Neimark, A. V., Adsorption-induced deformation of microporous carbons: Pore size distribution effect. *Langmuir* **2008,** 24, (13), 6603-6608.

22.Dolino, G.; Bellet, D.; Faivre, C., Adsorption strains in porous silicon. *Physical Review B* **1996,** 54, (24), 17919-17929.
23.Herman, T.; Day, J.; Beamish, J., Deformation of silica aerogel during fluid adsorption. *Phys. Rev. B* **2006,** 73, (9), 7.
24.Mushrif, S. H.; Rey, A. D., An integrated model for adsorption-induced strain in microporous solids *Chemical Engineering Science* **2009,** 64, (22), 4744-4753
25.Dourdain, S.; Britton, D. T.; Reichert, H.; Gibaud, A., Determination of the elastic modulus of mesoporous silica thin films by x-ray reflectivity via the capillary condensation of water. *Applied Physics Letters* **2009,** 93, (18), 183108.
26.Neimark, A. V.; Coudert, F. X.; Boutin, A.; Fuchs, A. H., Stress-Based Model for the Breathing of Metal−Organic Frameworks. *The Journal of Physical Chemistry Letters* **2010,** 1, (1), 445-449.
27.Lawnik, W. H.; Goepel, U. D.; Klauk, A. K.; Findenegg, G. H., Physisorption of cyclohexane on a SiO2/Si substrate : evidence of a wetting transition above the triple point. *Langmuir* **1995,** 11, (8), 3075-3082.
28.Muroyama, N.; Yoshimura, A.; Kubota, Y.; Miyasaka, K.; Ohsuna, T.; Ryoo, R.; Ravikovitch, P. I.; Neimark, A. V.; Takata, M.; Terasaki, O., Argon adsorption on MCM-41 mesoporous crystal studied by in situ synchrotron powder X-ray diffraction. *Journal of physical chemistry c* **2008,** 112, (29), 10803-10813.
29.Llewellyn, P. L.; Maurin, G.; Devic, T.; Loera-Serna, S.; Rosenbach, N.; Serre, C.; Bourrelly, S.; Horcajada, P.; Filinchuk, Y.; Ferey, G., Prediction of the Conditions for Breathing of Metal Organic Framework Materials Using a Combination of X-ray Powder Diffraction, Microcalorimetry, and Molecular Simulation. *J. Am. Chem. Soc.* **2008,** 130, 12808-12814.
30.Krasilnikova, O. K.; Bering, B. P.; Serpinskii, V. V.; Dubinin, M. M., Deformation of Zeolite CaNaX During Xenon Adsorption. *Bulletin of the Academy of Sciences of the USSR Division of Chemical Science* **1977,** 26, (5), 1099-1101.
31.Grosman, A.; Ortega, C., Influence of Elastic Strains on the Adsorption Process in Porous Materials: An Experimental Approach *Langmuir* **2009,** 25, (14), 8083-8093.
32.Eriksson, J. C., Thermodynamics of surface phase systems: V. Contribution to the thermodynamics of the solid-gas interface *Surface Science* **1969,** 14, (1), 221-246
33.Ash, S. G.; Everett, D. H.; Radke, C., Thermodynamics of Effects of Adsorption on Interparticle Forces. *Journal of the Chemical Society-Faraday Transactions II* **1973,** 69, (8), 1256-1277.
34.Scherer, G. W., Dilatation of Porous-Glass. *Journal of the American Ceramic Society* **1986,** 69, (6), 473-480.
35.Irving, J. H.; Kirkwood, J. G., The Statistical Mechanical Theory of Transport Processes. IV. The Equations of Hydrodynamics. *J. Chem. Phys.* **1950,** 18, 817.
36. Landau, L. D.; Lifshitz , E. M., *Theory of elasticity*. Butterworth-Heinemann 1986; p 195.
37.Biot, M., General theory of three dimensional consolidation. *Journal of Applied Physics* **1941,** 12, (2), 155-164.
38. Coussy, O., *Poromechanics*. John Wiley & Sons: 2004.
39.Jeffroy, M.; Fuchs, A. H.; Boutin, A., Structural changes in nanoporous solids due to fluid adsorption: thermodynamic analysis and Monte Carlo simulations. *Chem. Commun.* **2008**, 3275 - 3277.
40.Guenther, G.; Schoen, M., Sorption strains and their consequences for capillary condensation in nanoconfinement. *Molecular Simulations* **2009,** 35, (1-2), 138–150.
41.Ravikovitch, P. I.; Neimark, A. V., Density functional theory model of adsorption on amorphous and microporous silica materials. *Langmuir* **2006,** 22, (26), 11171-11179.
42.Mogilnikov, K. P.; Baklanov, M. R., Determination of Young's modulus of porous low-k films by ellipsometric porosimetry. *Electrochemical and Solid State Letters* **2002,** 5, (12), F29-F31.
43.Derjaguin, B. V., A theory of capillary condensation in the pores of sorbents and of other capillary phenomena taking into account the disjoining action of polymolecular liquid films. *Acta* *physicochimica URSS* **1940,** 12, 181.
44.Broekhoff, J. C. P.; De Boer, J. H., Studies on pore systems in catalysts : X. Calculations of pore distributions from the adsorption branch of nitrogen sorption isotherms in the case of open cylindrical pores B. Applications. *Journal of Catalysis* **1967,** 9, (1), 15-27.
45.Broekhoff, J. C. P.; de Boer, J. H., Studies on pore systems in catalysts : IX. Calculation of pore distributions from the adsorption branch of nitrogen sorption isotherms in the case of open cylindrical pores A. Fundamental equations. *Journal of Catalysis* **1967,** 9, (1), 8-14.
46.Neimark, A. V.; Ravikovitch, P. I., Capillary condensation in MMS and pore structure characterization. *Microporous and mesoporous materials* **2001,** 44, 697-707.
47.Ravikovitch, P. I.; Neimark, A. V., Density functional theory of adsorption in spherical cavities and pore size characterization of templated nanoporous silicas with cubic and three-dimensional hexagonal structures. *Langmuir* **2002,** 18, (5), 1550-1560.
48. Frenkel, J. I., *Kinetic theory of liquids*. Oxford University Press: London, 1946.
49.Dubinin, M. M.; Kataeva, L. I.; Ulin, V. I., *Bull. Acad. Sci. USSR, Div. Chem. Sci.* **1977,** 26, 459.
50.Jahnert, S.; Muter, D.; Prass, J.; Zickler, G. A.; Paris, O.; Findenegg, G. H., Pore Structure and Fluid Sorption in Ordered Mesoporous Silica. I. Experimental Study by in situ Small-Angle X-ray Scattering. *Journal of Physical Chemistry C* **2009,** 113, (34), 15201-15210.
51.Neimark, A. V.; Ravikovitch, P. I.; Vishnyakov, A., Bridging scales from molecular simulations to classical thermodynamics: density functional theory of capillary condensation in nanopores. *Journal of physics-condensed matter* **2003,** 15, (3), 347-365.
52.Derjaguin, B. V.; Churaev, N. V.; Muller, V. M., *Surface forces*. Consultants Bureau: New York, 1987; p 440.
53.Russo, P. A.; Carrott, M.; Padre-Eterno, A.; Carrott, P. J. M.; Ravikovitch, P. I.; Neimark, A. V., Interaction of water vapour at 298 K with Al-MCM-41 materials synthesised at room temperature. *Microporous and mesoporous materials* **2007,** 103, (1-3), 82-93.